\documentclass[aps,prb,reprint,superscriptaddress,showpacs,showkeys,preprintnumbers]{revtex4-1}

\usepackage[utf8]{inputenc}
\usepackage{graphicx}
\usepackage{bm}
\usepackage{amsmath}
\usepackage{amssymb}
\usepackage{hyperref}
\usepackage{subfigure}
\usepackage{float}
\usepackage{multirow}
\usepackage{booktabs}
\usepackage{varioref}
\usepackage{xr-hyper}
\usepackage[dvipsnames]{xcolor}
\usepackage{nicefrac}
\usepackage{xfrac}
\usepackage{hyperref}
\hypersetup{colorlinks,linkcolor=blue,urlcolor=blue,citecolor=blue}
\usepackage{ulem}
\usepackage{siunitx}
\usepackage{graphicx}
\usepackage{dcolumn}
\usepackage{braket}
\usepackage{wasysym}
\usepackage{textcomp}
\usepackage{upgreek}

\begin{document}

\author{Yujie~Yan}
\thanks{These authors contributed equally to this work.}
\affiliation{Department of Physics, The Chinese University of Hong Kong, Shatin, Hong Kong, China}

\author{Ying~Chan}
\thanks{These authors contributed equally to this work.}
\affiliation{Department of Physics, The Chinese University of Hong Kong, Shatin, Hong Kong, China}

\author{Xunyang~Hong}
\affiliation{Department of Physics, The Chinese University of Hong Kong, Shatin, Hong Kong, China}
\affiliation{Physik-Institut, Universit\"{a}t Z\"{u}rich, Winterthurerstrasse 190, CH-8057 Z\"{u}rich, Switzerland}

\author{S. Lin Er Chow}
\affiliation{Department of Physics, Faculty of Science, National University of Singapore, Singapore 117551, Singapore}

\author{Zhaoyang Luo}
\affiliation{Department of Physics, Faculty of Science, National University of Singapore, Singapore 117551, Singapore}

\author{Yuehong~Li}
\affiliation{Department of Physics, The Chinese University of Hong Kong, Shatin, Hong Kong, China}

\author{Tianren~Wang}
\affiliation{Department of Physics, The Chinese University of Hong Kong, Shatin, Hong Kong, China}

\author{Yuetong~Wu}
\affiliation{Department of Physics, The Chinese University of Hong Kong, Shatin, Hong Kong, China}

\author{Izabela~Bia\l{}o}
\affiliation{Physik-Institut, Universit\"{a}t Z\"{u}rich, Winterthurerstrasse 
190, CH-8057 Z\"{u}rich, Switzerland}

\author{Nurul Fitriyah}
\affiliation{Department of Physics, Faculty of Science, National University of Singapore, Singapore 117551, Singapore}

\author{Saurav Prakash}
\affiliation{Department of Physics, Faculty of Science, National University of Singapore, Singapore 117551, Singapore}

\author{Xing Gao}
\affiliation{Department of Physics, Faculty of Science, National University of Singapore, Singapore 117551, Singapore}

\author{King Yau Yip}
\affiliation{Department of Physics, Faculty of Science, National University of Singapore, Singapore 117551, Singapore}

\author{Qiang Gao}
\affiliation{Beijing National Laboratory for Condensed Matter Physics, Institute of Physics, Chinese Academy of Sciences, Beijing 100190, China}

\author{Xiaolin Ren}
\affiliation{Beijing National Laboratory for Condensed Matter Physics, Institute of Physics, Chinese Academy of Sciences, Beijing 100190, China}

\author{Jaewon~Choi}
\affiliation{Diamond Light Source, Harwell Campus, Didcot, Oxfordshire OX11 0DE, United Kingdom}
\affiliation{Department of Physics, Korea Advanced Institute of Science and Technology, 291 Daehak-ro, Daejeon 34141, Republic of Korea}

\author{Ganesha Channagowdra}
\affiliation{National Synchrotron Radiation Research Center, Hsinchu 300092, Taiwan}

\author{Jun~Okamoto}
\affiliation{National Synchrotron Radiation Research Center, Hsinchu 300092, Taiwan}

\author{Xingjiang Zhou}
\affiliation{Beijing National Laboratory for Condensed Matter Physics, Institute of Physics, Chinese Academy of Sciences, Beijing 100190, China}

\author{Zhihai Zhu}
\affiliation{Beijing National Laboratory for Condensed Matter Physics, Institute of Physics, Chinese Academy of Sciences, Beijing 100190, China}

\author{Liang~Si}
\affiliation{School of Physics, Northwest University, Xi’an 710127, China}

\author{Mirian~Garcia-Fernandez}
\affiliation{Diamond Light Source, Harwell Campus, Didcot, Oxfordshire OX11 0DE, United Kingdom}

\author{Ke-Jin~Zhou}
\affiliation{Diamond Light Source, Harwell Campus, Didcot, Oxfordshire OX11 0DE, United Kingdom}

\author{Hsiao-Yu Huang}
\affiliation{National Synchrotron Radiation Research Center, Hsinchu 300092, Taiwan}

\author{Di-Jing Huang}
\affiliation{National Synchrotron Radiation Research Center, Hsinchu 300092, Taiwan}

\author{Johan~Chang}
\email{johan.chang@physik.uzh.ch}
\affiliation{Physik-Institut, Universit\"{a}t Z\"{u}rich, Winterthurerstrasse 190, CH-8057 Z\"{u}rich, Switzerland}

\author{A.~Ariando}
\email{ariando@nus.edu.sg}
\affiliation{Department of Physics, Faculty of Science, National University of Singapore, Singapore 117551, Singapore}

\author{Qisi~Wang}
\email{qwang@cuhk.edu.hk}
\affiliation{Department of Physics, The Chinese University of Hong Kong, Shatin, Hong Kong, China}

%\title{Persistent paramagnon in Sm-based infinite-layer nickelate superconductors}

\title{Persistent paramagnons in high-temperature infinite-layer nickelate superconductors}

\maketitle

\noindent\textbf{
The recent discovery of high-temperature superconductivity in hole-doped SmNiO$_2$, exhibiting the record-high transition temperature $T_c$ among infinite-layer (IL) nickelates, has opened a new avenue for exploring design principles of superconductivity.
Experimentally determining the electronic structure and magnetic interactions in this new system is crucial to elucidating the mechanism behind the enhanced superconductivity. Here, we report a Ni $L$-edge resonant inelastic x-ray scattering (RIXS) study of superconducting Sm-based IL nickelate thin films Sm$_{1-x-y-z}$Eu$_x$Ca$_y$Sr$_z$NiO$_2$ (SECS). Dispersive paramagnonic excitations are observed in both optimally and overdoped SECS samples, supporting a spin-fluctuation-mediated pairing scenario. However, despite the two-fold enhancement of $T_c$ in the Sm-based nickelates compared to their Pr-based counterparts, the effective exchange coupling strength is reduced by approximately $20$\%. This behavior contrasts with hole-doped cuprates, where magnetic interactions correlate positively with $T_c$, highlighting essential differences in their superconducting mechanisms.}\\

\noindent The discovery of superconductivity in nickelate materials---featuring both square-planar and Ruddlesden-Popper (RP) structures---has unveiled a new family of unconventional high-temperature superconductors~\cite{li_superconductivity_2019,pan_superconductivity_2021,sun_signatures_2023,zhu_superconductivity_2024}. With a similar crystal structure and nominal $d^9$ electronic configuration, the infinite-layer (IL) nickelates have been proposed as analogs to the cuprate superconductors~\cite{anisimov_electronic_1999,botana_similarities_2020}. In fact, similar antiferromagnetic excitations and a hole-like Fermi pocket with $3d_{x^2-y^2}$ character are observed in IL-nickelate and cuprate superconductors~\cite{lu_magnetic_2021,ding_cuprate_2024,sun_electronic_2025,li_observation_2025}. Strange metal behaviour around optimally doping is another commonality~\cite{lee_linear_2023,hsu_transport_2024}.
However, distinctions in their electronic behaviors have also been revealed. For example, while the parent cuprates are considered as charge-transfer insulators, the IL nickelates are closer to the Mott-Hubbard regime, due to reduced ligand-oxygen hybridization~\cite{lee_infinite_2004,botana_similarities_2020,kitatani_nickelate_2020,chen_electronic_2022}. 
In addition to the correlated cuprate-like hole pocket from Ni $d_{x^2-y^2}$ band, a three-dimensional electron pocket is observed at the Brillouin zone corner~\cite{ding_cuprate_2024,sun_electronic_2025,li_observation_2025}, though the role of multi-orbital physics remains uncertain~\cite{si_closing_2024,lechermann_multiorbital_2020}.
Moreover, while hole-doped cuprates exhibit clear signatures of charge inhomogeneity~\cite{comin_resonant_2016,von_arx_fate_2023}, the existence of charge order in IL nickealtes is still controversial~\cite{rossi_broken_2022,tam_charge_2022_nm,krieger_charge_2022,ren_two_2024,parzyck_absence_2024,oppliger_discovery_2025}.

Another contrast is the role played by the rare-earth ions in determining the low-energy physical properties. Whereas rare-earth elements have a marginal influence on the electronic structure of cuprates, significant hybridization between rare-earth $5d$ and Ni $3d$ bands has been observed in IL nickelates~\cite{hepting_electronic_2020}. 
Self-doping from the rare-earth $5d$ states into the NiO$_2$ planes has presumably led to a metallic state and the absence of magnetic order in the parent compounds~\cite{li_superconductivity_2019,li_superconducting_2020,lee_linear_2023}.
Experiments further revealed distinct electronic behaviors among IL nickelates with different rare-earth elements~\cite{harvey_evidence_2022,chow_pairing_2022,chow_pauli_2022,wang_effects_2023}. For example, while Sr-doped LaNiO$_{2}$ and PrNiO$_{2}$ exhibit strongly anisotropic upper critical field $H_{c2}$~\cite{wang_effects_2023}, an unexpected isotropic $H_{c2}$ is revealed in (Nd,Sr)NiO$_{2}$~\cite{wang_isotropic_2021,wang_effects_2023}. 
However, the origin of these differences, and their implications for superconductivity, remain subjects of intense debate~\cite{rossi_universal_2024,been_electronic_2021,zhang_rare-earth_2023}.

\begin{figure*}[t]
    \centering
 \includegraphics[width=.95\textwidth]{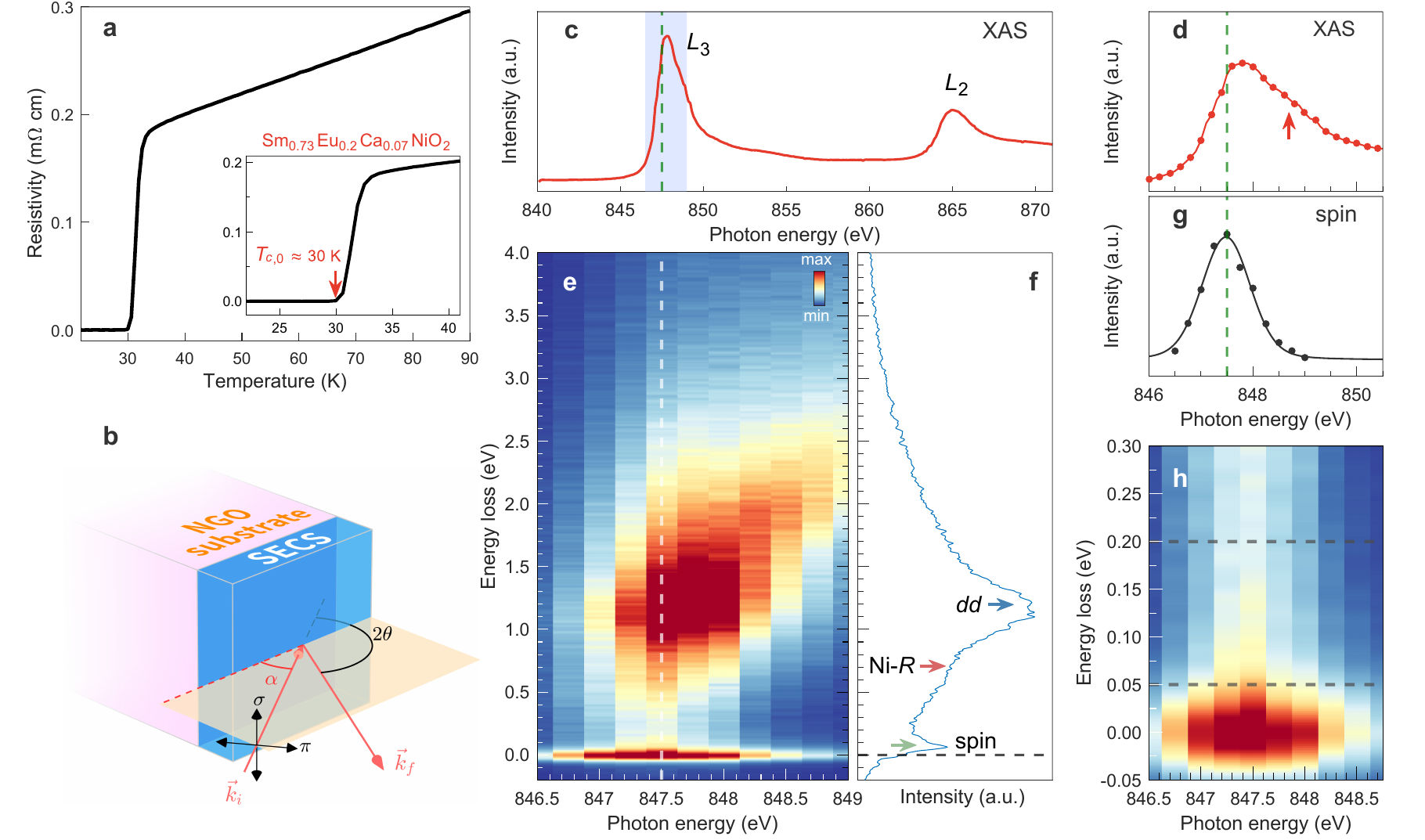}

    \caption{Resistivity measurement, XAS, and RIXS intensity map for optimally doped SECS. (a) The temperature-dependent resistivity for the OP SECS film used for RIXS measurements. (b) Scattering geometry of the RIXS experiment. The wave vectors of the incident and scattered photons ($\vec{k}_i$ and $\vec{k}_f$) define the scattering angle $2\theta$. (c) XAS spectrum across the Ni $L_3$ ($\sim$$847.5$~eV) and $L_2$ ($\sim$$865$~eV) resonances measured at $30$~K using total electron yield. (d) XAS across the Ni $L_3$ edge. An additional shoulder (red arrow) is observed at $\sim$$1$~eV above the main peak. (e) RIXS intensity as a function of incident photon energy and energy loss, measured at $30$~K with an incident angle $\alpha=118.3^{\circ}$. The range of incident energy is indicated by the blue-shaded area in (c). (f) RIXS spectrum taken at the resonance energy (green dashed lines in (c,d,g); white dashed line in (e)). Elastic scattering has been subtracted to enhance the visibility of low-energy features. Blue, red and green arrows indicate excitations around $1.2$ eV, $0.7$ eV and $0.1$ eV, respectively. (g) Resonant profile of the $\sim$$0.1$~eV excitation (intensity integrated between $0.05$-$0.2$ eV). (h) Enlarged view of the low-energy region in (e). Gray dashed lines mark the intensity integration range for (g). a.u., arbitrary units.}
    \label{fig:fig1}
\end{figure*}

Recently, superconductivity was realized in high-crystallinity, phase-pure, hole-doped Sm-based IL nickelate thin films~\cite{chow_bulk_2025,yang_enhanced_2025}, which exhibit an optimal superconducting onset temperature $T_c$ approaching $40$~K with zero-resistance state at $T_{c,0}\approx30$~K---the highest reported among all IL nickelates to date. 
The nearly two-fold increase in $T_c$ compared to hole-doped (Nd/La/Pr)NiO$_2$ 
systems highlights Sm-based compounds as a promising platform for understanding the mechanism of superconductivity in these materials, especially the role of the rare-earth elements.
While some studies suggest that reduced interlayer spacing~\cite{yang_enhanced_2025,wang_pressure-induced_2022} or smaller in-plane lattice constants~\cite{ren_possible_2023,lee_synthesis_2025} may enhance $T_c$, the associated changes in electronic structure and magnetic interactions remain unclear---yet are essential for uncovering the microscopic mechanism of superconductivity.

In this study, we employ resonant inelastic x-ray scattering (RIXS) at Ni $L$-edge to investigate the electronic and magnetic excitations in superconducting Sm$_{1-x-y-z}$Eu$_x$Ca$_y$Sr$_z$NiO$_2$ (SECS) thin films.
An optimally doped (OP) sample Sm$_{0.73}$Eu$_{0.2}$Ca$_{0.07}$NiO$_2$ (nominal doping $p=0.19$) with a superconducting onset temperature $T_{c,\rm{onset}} \approx 35$~K (Fig.~\ref{fig:fig1}a), and a non-superconducting overdoped (OD) sample Sm$_{0.53}$Eu$_{0.4}$Ca$_{0.07}$NiO$_2$ (nominal doping $p=0.31$) are studied (Supplementary Note~1 and ref.~\onlinecite{chow_bulk_2025}). 
To distill the key ingredients for superconductivity in IL nickelates, comparative measurements are performed on an optimally doped Pr$_{0.8}$Sr$_{0.2}$NiO$_2$ (PSNO) thin film with $T_{c,\rm{onset}} \approx 9$~K~\cite{ren_possible_2023}.
We observe dispersive paramagnonic excitations in SECS, with a bandwidth of approximately 100 meV, reduced by $\sim$$20$\% compared to PSNO.
The persistence of robust magnetic correlations with large exchange couplings across these IL nickelate systems supports the scenario in which superconducting pairing is primarily mediated by spin fluctuations~\cite{worm_spin_2024,katukuri_electronic_2020,sakakibara_model_2020,wu_robust_2020,lane_competing_2023,qin_intrinsic_2025}.
Nevertheless, the significantly enhanced $T_c$ in Sm-based nickelates deviates from the cuprate-like trend, where larger exchange couplings correlate positively with higher $T_c$. Our findings thus suggest that additional factors, such as the multi-band electronic structure, enhanced three-dimensionality, and Kondo-like hybridization involving rare-earth orbitals, need to be considered to fully capture the microscopic mechanism underlying superconductivity in IL nickelates.\\

\begin{figure*}[t]%[thb]
    \centering
    \includegraphics[width=0.98\textwidth]{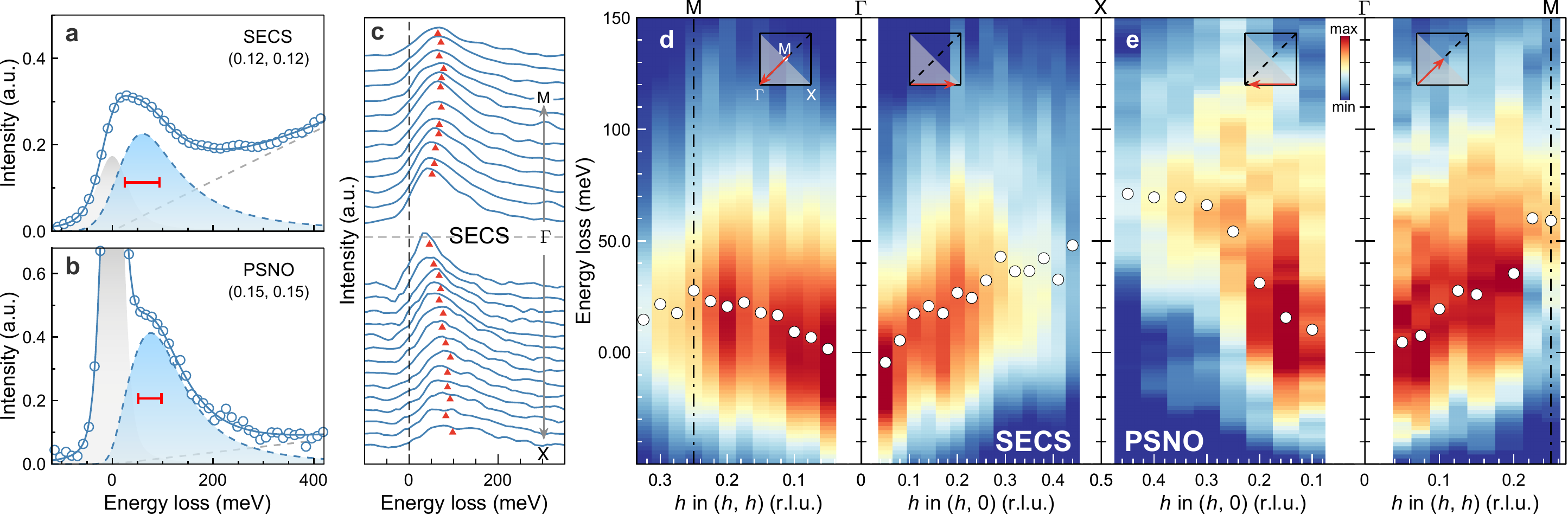}
    \caption{Paramagnonic excitations in optimally doped SECS and PSNO. (a,b) Representative raw RIXS spectra of OP (a) SECS and (b) PNSO samples.
    The blue fitting curve consists of three components: elastic scattering (gray), paramagnon excitation (light blue), and high-energy  background (gray dashed line). The red horizontal bars denote the instrumental energy resolution. 
    To reduce elastic scattering, a scattering angle $2\theta=90^{\circ}$ is used in (a).
    (c) Momentum-dependent RIXS spectra of OP SECS. Elastic and background scattering has been subtracted. Red arrows indicate the position of the maximum intensity of the fitted paramagnonic excitations. (d,e) RIXS intensity map measured along momentum directions $(h, 0)$ and $(h, h)$ in OP (d) SECS and (e) PSNO. White dots indicate the peak energy ($\omega_{\text{max}}$) of the magnetic excitation mode extracted from fits (see main text).
    Black dash-dot lines mark the antiferromagnetic zone boundary. The scan directions of the in-plane momentum are illustrated in the insets. a.u., arbitrary units.}
    \label{fig:fig2}
\end{figure*}

\noindent\textbf{Results}\\
X-ray absorption spectrum (XAS) of the optimally doped SECS thin film across the Ni $L_3$ and $L_2$ edges is shown in Fig.~\ref{fig:fig1}c.
A shoulder---due to doped holes in the Ni $d_{x^2 - y^2}$ orbital~\cite{rossi_orbital_2021}---appears about $1$~eV above the Ni $L_3$ peak (Fig.~\ref{fig:fig1}d).
To probe the electronic excitations, we performed RIXS measurements with the incident photon energy varied across the Ni $L_3$ edge---indicated by the blue-shaded area in Fig.~\ref{fig:fig1}c.
Fluorescence dominated RIXS intensities are observed at energies of $2.5$~eV and above (see Fig.~\ref{fig:fig1}e).
The $dd$ excitations centered around $1.2$~eV resemble those observed in hole-doped Nd$_{1-x}$Sr$_x$NiO$_2$~\cite{rossi_orbital_2021}, indicating a similar Ni crystal field environment.
The excitation at $\sim$$0.7$~eV (Fig.~\ref{fig:fig1}f) is attributed to hybridization between Ni $3d$ and rare-earth $5d$ orbitals~\cite{hepting_electronic_2020}, evidencing occupied rare-earth 5$d$ states below the Fermi level in the superconducting regime.
A low-energy excitation is identified around $0.1$~eV, which is most prominent at an incident energy of 847.5~eV (Fig.~\ref{fig:fig1}h).
The resonant behavior demonstrates its electronic origin (Fig.~\ref{fig:fig1}g,h).

To establish the nature of this low-energy excitation, we performed RIXS measurements (Fig.~\ref{fig:fig1}b) along the high-symmetry $(h,0)$ and $(h,h)$ momentum directions at the Ni $L_3$ edge ($847.5$~eV). Each spectrum is fitted with a three-component model---see Fig.~\ref{fig:fig2}a,b. The quasi-elastic scattering is described by a Voigt profile (shown in gray); the low-energy excitation mode is modeled as a damped harmonic oscillator (DHO) convoluted with the instrumental resolution (blue shaded area); the background, primarily arising from the particle-hole continuum, is approximated by a quadratic function (gray dashed line). 
In this fashion, we extract the momentum dependence of the low-energy mode, as summarized in Fig.~\ref{fig:fig2}c,d.

With increasing momentum---along $(h,0)$ and $(h,h)$ directions---the excitation mode shifts to higher energy (Fig.~\ref{fig:fig2}c,d), displaying the characteristic dispersion of a square-lattice antiferromagnetic paramagnon. This is also consistent with previous observations of propagative spin excitations in other IL nickelates~\cite{lu_magnetic_2021,rossi_universal_2024,gao_magnetic_2024}. 
The reduced zone-boundary energy at $(0.25,0.25)$ compared to $(0.5,0)$ reveals the existence of higher-order exchange interactions~\cite{wang_magnon_2024,bialo_strain-tuned_2024}.
The broad energy width further supports our assignment of this excitation as a paramagnon. As the in-plane momentum transfer increases, the spectral weight is suppressed, and the energy linewidth broadens.

\begin{figure*}[thb]
    \centering
 \includegraphics[width=.9\textwidth]{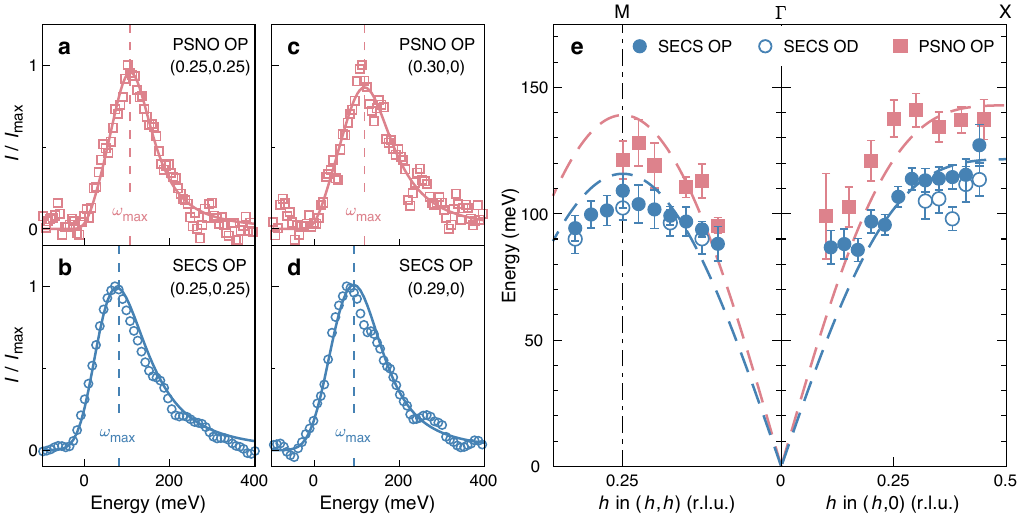}
    \caption{Dispersions of paramagnon in SECS and PSNO. 
    (a-d) Paramagnon spectral components for OP PSNO (red) and SECS (blue) near the antiferromagnetic zone boundaries in the $(h,h)$ (a,b) and $(h,0)$ (c,d) directions. Dashed lines indicate the energy positions of the peak maxima (${\omega}_\text{max}$). The intensity $I$ is normalized to its peak value $I\rm_{max}$.
    (e) Paramagnon pole energies ($\omega_0$) extracted from fits to the RIXS spectra for SECS (blue) and PSNO (red). Error bars represent the fitting uncertainties. Dashed lines denote fits to an effective Heisenberg model (see main text) for OP SECS (blue) and PSNO (red).}
    \label{fig:fig3}
\end{figure*}

The paramagnon persists in the overdoped SECS sample, with both the pole energy $\omega_0$ and the peak energy $\omega_{\text{max}}$ decreasing with increased doping (see Fig.~\ref{fig:fig3}e and Supplementary Fig.~1). This behavior resembles the doping evolution of paramagnon excitations observed in hole-doped cuprate superconductors~\cite{meyers_doping_2017}, reflecting a gradual softening of spin correlations as additional carriers are introduced.\\

\noindent\textbf{Discussion}\\
To gain insight into the influence of rare-earth element on the magnetism, we compare the paramagnon in OP SECS and PSNO. As shown in Fig.~\ref{fig:fig2}d and e, the paramagnon in SECS display a comparable but slightly smaller bandwidth. 
The effective exchange interactions are derived from the paramagnon pole ($\omega_0$) using a linear spin-wave model (see Supplementary Information for details). Given the weak but discernible zone-boundary dispersion, we included both nearest-neighbor ($J_1$) and next-nearest-neighbor ($J_2$) exchange couplings:

\begin{equation}
H = J_1 \sum_{\langle i, j \rangle} \mathbf{S}_i \cdot \mathbf{S}_j + J_2 \sum_{\langle i, i' \rangle} \mathbf{S}_i \cdot \mathbf{S}_{i'}
\end{equation}
where $\langle i, j \rangle$ and $\langle i, i' \rangle$ denote pairs of nearest and next-nearest neighbor spin sites, respectively. The best fit yields $J_1=46.6\pm4.9$~meV and $J_2= -2.4\pm3.7$~meV for OP SECS, and $J_1=57.3\pm7.1$~meV and $J_2= -1.6\pm5.2$~meV for OP PSNO, assuming spin $S=1/2$ and quantum renormalization factor $Z_c=1.18$~\cite{lu_magnetic_2021,wang_magnon_2024}.
Applying the same analysis to the paramagnon dispersion for Nd$_{0.775}$Sr$_{0.225}$NiO$_2$~\cite{lu_magnetic_2021}, we find a consistently larger magnon bandwidth and stronger exchange coupling compared to OP SECS (see Supplementary Fig.~2 and Supplementary Table~1).

Due to the difficulties in achieving phase-pure undoped SmNiO$_2$ thin film~\cite{raji_valence-ordered_2024}, we studied optimally and overdoped SECS samples free from extended defects. Hole doping into the half-filled state effectively dilutes the spin sites. As in the cuprates~\cite{monney_resonant_2016,le_tacon_dispersive_2013}, this leads to damping and softening of the magnon energy, and thus reduced effective exchange couplings compared to the undoped compounds~\cite{lu_magnetic_2021,gao_magnetic_2024,rossi_universal_2024}.

A $\sim$$20$\% smaller in-plane exchange $J_{\parallel}$ in SmNiO$_2$ compared to (Nd/La/Pr)NiO$_2$ has been theoretically predicted by first-principles calculations~\cite{zhang_rare-earth_2023}. It is suggested that smaller rare-earth ionic radii enhance the three-dimensional character of the electronic structure. While the exchange interactions remain predominately mediated by the Ni $d_{x^2-y^2}$ orbitals, the increased three-dimensionality, accompanied by subtle structural modifications~\cite{zhang_rare-earth_2023,subedi_possible_2023}, may introduce a finite out-of-plane exchange coupling $J_\perp$. In fact, recent ARPES experiments have observed additional electron Fermi pockets in hole-doped LaNiO$_{2}$~\cite{ding_cuprate_2024,sun_electronic_2025} and NdNiO$_{2}$~\cite{li_observation_2025}. Future experiments will be essential to clarify how the electronic structure evolves across different rare-earth IL nickelates.

The observation of large exchange couplings across all IL nickelate systems is consistent with theoretical proposals that spin fluctuations play a dominant role in the superconducting pairing~\cite{worm_spin_2024,katukuri_electronic_2020,sakakibara_model_2020,wu_robust_2020,lane_competing_2023,qin_intrinsic_2025}. Yet, the reduced in-plane exchange coupling $J_{\parallel}$ in SECS despite its enhanced $T_c$ is markedly different from cuprates, where a positive correlation has been established between $T_c$ and the paramagnon energy scale~\cite{wang_paramagnons_2022,ivashko_strain-engineering_2019}. This difference could possibly originate from the multi-band nature of nickelates. Compared to cuprates, the hybridization between the rare-earth 5$d$ and Ni 3$d$ states in the IL nickelates renders them relevant to the low-energy physics.
The enhanced three-dimensionality may also strengthen Kondo-like coupling effects~\cite{zhang_self-doped_2020,wang_distinct_2020,lechermann_multiorbital_2020}.
Our findings suggest that these differences between IL nickelates and cuprates could be fundamental for understanding their superconducting mechanisms. 
Furthermore, the nontrivial relationship between in-plane exchange coupling and $T_c$ revealed by our study points toward alternative design principles for achieving higher $T_c$ in nickelate superconductors.\\

\noindent\textbf{Methods}\\
\noindent\textbf{Film growth}\\
\noindent The precursor thin films of Sm$_{1-x-y-z}$Eu$_x$Ca$_y$Sr$_z$NiO$_2$ (SECS, $6$~nm-thick) and Pr$_{0.8}$Sr$_{0.2}$NiO$_2$ (PSNO, $8$~nm-thick) were grown on NdGaO$_3$(110) and SrTiO$_3$(001) substrates, respectively, using pulsed laser deposition (PLD), followed by a topotactic reduction process described in refs.~\onlinecite{chow_bulk_2025,ren_possible_2023}. The SECS and PSNO films were capped with SrTiO$_3$ layers of $1$~nm and $14$~nm thickness, respectively.
The optimal $T_{c,0}\approx30$~K for SECS is achieved at a hole doping level of $p\approx0.17\text{-}0.19$~\cite{chow_bulk_2025}.\\

\noindent\textbf{RIXS experiments}\\
\noindent Ni $L$-edge RIXS experiments on SECS and PNSO were carried out at the 41A1 beamline~\cite{singh_development_2021} at the Taiwan Photon Source and I21 beamline~\cite{zhou_i21_2022} at the Diamond Light Source, respectively.
The wave vector $\mathbf{Q}$ was defined as $(h,k,l)=(ha/2\pi,kb/2\pi,lc/2\pi)$ in reciprocal lattice units (r.l.u.), with $a=b=3.86$~\AA~and $c=3.27$~\AA~ for SECS~\cite{chow_bulk_2025}, and $a=b=3.905$~\AA~and $c=3.37$~\AA~for PSNO~\cite{osada_phase_2020}.
To maximize the accessible range of in-plane momentum transfer, the 41A1 (I21) spectrometer
was positioned at the largest scattering angle of $2\theta=150^{\circ}~(154^{\circ})$ unless otherwise indicated, and the sample temperature was set to $30~(16)$~K. The combined energy resolution $\delta$, characterized by the full-width-at-half-maximum (FWHM) of the elastic scattering profile of amorphous carbon, was $\delta=67$~meV ($46$~meV) for measurements at 41A1 (I21). The incident x-rays are $\pi$-polarized to enhance the single spin-flip cross section. The beamline-specific photon energy offset has not been considered.
RIXS intensities were normalized to the weight of the $dd$ excitations between 0.5~eV and 4~eV, as in refs.~\onlinecite{ghiringhelli_long-range_2012,lin_strongly_2020,wang_charge_2021,arpaia_signature_2023}.\\

\noindent\textbf{Fitting of RIXS spectra}\\
\noindent Elastic scattering and paramagnon components are fitted using a Voigt and a damped harmonic oscillator function, respectively. The width of the elastic profile is set by the instrumental energy resolution $\delta$ and the damped harmonic oscillator

\begin{equation}
\chi''(\mathbf{Q}, \omega) = \frac{\chi'(\mathbf{Q}) \, \gamma(\mathbf{Q}) \, \omega}{\left( \omega^2 - \omega_{0}(\mathbf{Q})^2 \right)^2 + \omega^2 \gamma(\mathbf{Q})^2}
\end{equation}
is Gaussian-convoluted with the resolution $\delta$, and multiplied by the thermal factor (see Supplementary Information for details). Here, $\gamma$ and $\omega_0$ represent the damping factor and the paramagnon pole, respectively. $\chi'$ is the real part of the zero frequency susceptibility.\\

\noindent\textbf{Acknowledgments}\\
We thank Karsten Held and Hanghui Chen for helpful discussions. The work at CUHK is supported by the Research Grants Council of Hong Kong (ECS No. 24306223), and the Guangdong Provincial Quantum Science Strategic Initiative (GDZX2401012). X.H., I.B. and J.C. thank the Swiss National Science Foundation under Projects No. 200021\_188564. The work at NUS is supported by the Ministry of Education (MOE), Singapore, under its Tier-2 Academic Research Fund (AcRF), Grant No. MOE-T2EP50123-0013, by the SUSTech-NUS Joint Research Program, and by the MOE Tier-3 Grant (MOE-MOET32023-0003) `Quantum Geometric Advantage'. 
Part of this research from IOP is supported by the National Key Research and Development Program of China (Grant No.  2021YFA1401800 and 2022YFA1403900 ).
We acknowledge the 41A1 and I21 Beamlines for providing beamtime under Proposals 2025-1-229-1 and MM30189.\\

\noindent\textbf{Author contributions}\\
Q.W. conceived the project. S.L.E.C., N.F., S.P., X.G., Z.L. and K.Y.Y. grew and characterized the SECS films supervised by A.A. Q.G., X.R., X.Z. and Z.Z. grew the PSNO film. 
Y.Y., Y.C., Y.L., T.W., Y.W., G.C., J.O., H.H., D.H. and Q.W. carried out the RIXS experiments at 41A1. I.B., J.Choi, M.G.F., K.J.Z., J.Chang and Q.W. carried out the RIXS experiments at I21. Y.Y. and X.H. analyzed the RIXS data with the help from L.S. J.Chang, A.A. and Q.W. wrote the manuscript with input from other authors.\\

\noindent\textbf{Data availability} \\
Data supporting the findings of this study are available from the corresponding authors on a reasonable request.\\

\noindent\textbf{Competing interests} \\
The authors declare no competing interests.\\

\end{document}